\documentclass[useAMS,usenatbib,usegraphicx]{mn2e}

\title[]{Kinematics of the H$_2$O masers at the centre of the PN K3-35}
\author[Uscanga et al.]{L. Uscanga,$^{1,2}$\thanks{E-mail:
lucero@iaa.es} Y. G\'omez,$^{1}$ A. C. Raga,$^{3}$ 
J. Cant\'o,$^{4}$ G. Anglada,$^{2}$ 
\newauthor
J. F. G\'omez $^{2}$, J. M. Torrelles,$^{5}$ and L. F. Miranda $^{2}$\\
$^{1}$Centro de Radioastronom\'{\i}a y Astrof\'{\i}sica, UNAM, Apartado Postal 3-72, 58089 Morelia, Michoac\'an, Mexico\\
$^{2}$Instituto de Astrof\'{\i}sica de Andaluc\'{\i}a, CSIC, Apartado 3004, E-18080 Granada, Spain\\
$^{3}$Instituto de Ciencias Nucleares, UNAM, Apartado Postal 70-543, 04510 M\'exico, DF, Mexico\\
$^{4}$Instituto de Astronom\'{\i}a, UNAM, Apartado Postal 70-264, 04510 M\'exico, DF, Mexico\\
$^{5}$Instituto de Ciencias del Espacio (CSIC)-IEEC, Facultat de F\'{\i}sica, Universitat de Barcelona, E-08028 Barcelona, Spain}
\begin{document}


\pagerange{\pageref{firstpage}--\pageref{lastpage}} \pubyear{2008}

\maketitle

\label{firstpage}

\begin{abstract}
We have studied the kinematics traced by the water masers located at the centre of the planetary nebula (PN) K3-35, using data from previous Very Large Array (VLA) observations.
An analysis of the spatial distribution and line-of-sight velocities of the maser spots allows us to identify typical patterns of a rotating and expanding ring in the position-velocity diagrams, according to our kinematical model. We find that the distribution of the masers is compatible with tracing a circular ring with a $\simeq$0\farcs021 ($\simeq$100~AU) radius, observed with an inclination angle with respect to the line of sight of 55$\degr$. 
We derive expansion and rotation velocities of 1.4 and 3.1 km s$^{-1}$, respectively. The orientation of the ring projected on the plane of the sky, at PA$\simeq$158$\degr$, is almost orthogonal to the direction of the innermost region of the jet observed in K3-35, suggesting the presence of a disc or torus that may be related to the collimation of the outflow. 
\end{abstract}

\begin{keywords}
masers -- planetary nebulae: general -- planetary nebulae: individual: K3-35
\end{keywords}

\section{Introduction}
High angular resolution observations of molecular gas have revealed the presence of dense equatorial discs and tori towards several late asymptotic giant branch (AGB) stars and young planetary nebulae (PNe): for instance, \citet{bie88}, \citet{for98}, \citet{sah98b}, \citet{buj03}. 
The interaction between the post-AGB wind and such equatorial structures
has been proposed as one of the possible physical mechanisms in shaping the bipolar and multipolar morphologies seen in PNe and proto-PNe (PPNe) \citep{bal87,mel95,sok00,ick03}. 

The origin of molecular discs and tori around late AGB stars is not completely clear, but a possible explanation is the presence of a binary system \citep{mor87,liv88,taa89,sok06}. In this case, when one of the stars enters the AGB phase, some of the ejected material can be retained, generating an extended torus. However, in the case of the high velocity jets observed in some late AGB stars, post-AGB stars, and young PNe \citep{fei85,sah98a,ima02,rie03} it is believed that a more effective collimation mechanism(s) should be present in the innermost region of the object, such as an accretion disc \citep{mor87,sok94,sok00} or a stellar magnetic field produced by a rotating star \citep{pas85,che94,gar97}. 
Furthermore, recent observations reveal compelling evidence for magnetic fields in post-AGB stars and PNe, associated with equatorial discs and/or jets (e.g. \citealt{sab07,vle06}).					
			
The process for the formation of accretion discs in PPNe has been investigated in detail by \citet{rey99}. They find that a disc forms when a close binary system (with a substellar companion) undergoes common envelope evolution. 
For the case in which a low-mass secondary is disrupted during a dynamically unstable mass transfer process, an accretion disc, with a radius of $\sim$10~AU and a mass of $\sim$2$\times10^{-3}$~M$_{\sun}$, forms within $\sim$100~yr.
Recently, \citet{rij05} from their 3D simulations based on a two-wind model with a warped disc, suggested that, in order to explain the observed multipolar and point-symmetric shape of PNe, the required discs are quite small ($\sim$10--100~AU). Furthermore, these disc-like structures should be dense ($10^7-10^8$~cm$^{-3}$) and in Keplerian rotation.

Interferometric CO observations show larger toroidal molecular structures with sizes in the range of $\simeq$1000--6000 AU in PPNe and young PNe.
These structures seem to be systematically
in expansion with a mean velocity of $\sim$7~km~s$^{-1}$, such as in M~1-92 \citep{buj98}, M~2-9 \citep{zwe97}, \mbox{M~2-56} \citep{cas02}, or KjPn~8 \citep{for98}. These velocities are comparable to or below those found in expanding AGB envelopes \citep{hug07}. 
Importantly, rotation has been observed in the Red Rectangle as well as slower expansion ($\sim$0.8~km~s$^{-1}$), superimposed on rotation, according to \citet{buj05}.
Note that the sizes of the tori are about one or two orders of magnitude more than the sizes of the disc-like structures proposed by \citet{rey99} and \citet{rij05}.

Maps of water maser emission allow the identification of a disc-like structure with a radius of 0\farcs12 in IRAS 17347-3139 \citep{deg04}, which corresponds to $\simeq$100--750~AU at the source distance \citep{gom05a}. In this case, the gas kinematics suggests the presence of both rotation and expansion in the disc traced by the water masers.
In this context, it is very important to study in detail the kinematics of much smaller disc-like structures that might be related to the collimation of the observed bipolar outflows in some PNe.

K3-35 is a young PN that shows a bipolar outflow in optical images \citep{mir00}.
At radio wavelengths, K3-35 exhibits a bright core and two bipolar lobes with an S-shape \citep{mir01}.
The distance to this object has been estimated to be $\sim$5~kpc \citep{zha95}, using an statistical method. However, we note the large uncertainty of this type of estimate, since the application of different statistical methods could give distances varying by factors of $\sim$3 \citep{phi04}. 
The characteristic S-shape morphology of the radio lobes can be successfully reproduced by a precessing jet evolving in a dense circumstellar medium \citep{vel07}. Water maser emission has been found in three regions: two regions located at the tips of the bipolar radio jet about $\simeq$1$''$
from the centre (regions N and S), and another region towards the core of the nebula within $\simeq$0\farcs02 (region C), suggesting the presence of a torus \citep{mir01}. 
In addition, OH maser emission has been detected towards the centre of K3-35 (within $\simeq$0\farcs04), showing circular polarisation that suggests the presence of a magnetic field \citep{mir01,gom05b}. 

We decided to study the spatio-kinematical distribution of the water masers reported by \citet{mir01} towards the centre of K3-35 in order to identify possible expansion and/or rotation motions. 
The paper is organised as follows. 
In Section 2, we present a simple kinematical model of a ring including both expansion and rotation, and we calculate the pattern delineated in the position-velocity diagrams. We also describe the least-squares fit procedure that we used. In Section 3, we present the current observational data of H$_2$O masers in the PN K3-35. We then apply our model to the H$_2$O masers located towards the centre of the PN K3-35, making a comparison between the results and the observations. Finally, in Section 4, we discuss the implications of our results.

\section[]{Rotating and Expanding Ring Model}
\subsection{Model}
We assume a narrow, uniform, rotating, and expanding ring of radius $R$, arbitrarily oriented with respect to the line of sight. Its projection on the plane of the sky is an ellipse with semimajor and semiminor axes $a$ and $b$, respectively.
We define the two frames of reference shown in Fig.~1. Both coincide with the plane of the sky, one of them has its origin at the centre of the ellipse and is oriented such that the $x'$-axis is along the major axis of the projected ellipse, and the other one has the axes parallel to the RA and Dec axes.
The semimajor and semiminor axes are related to the ring radius by $a=R$ and $b=R\cos i$, where $i$ is the inclination angle between the line of sight and the normal to the ring plane as shown in Fig.~2.

The equation of the ellipse is given by
\begin{equation}
   \frac{x'^2}{a^2}+\frac{y'^2}{b^2}=1,
\end{equation}
and the transformation equations between the coordinate systems are
\begin{equation}
x'=(x-x_0)\cos\theta+(y-y_0)\sin\theta, 
\end{equation}
\begin{equation}
y'=-(x-x_0)\sin\theta+(y-y_0)\cos\theta,
\end{equation}
 where
 ($x_0,y_0$) is the position of the centre of the ellipse, and
 $\theta$ is the angle between the $x$-axes of the two frames of reference, and is defined as positive clockwise. The
angle $\theta$ is related to the position angle (PA) of the major axis of the ellipse by $\rmn{PA}=90\degr-\theta$.

\begin{figure}
\includegraphics[width=72mm]{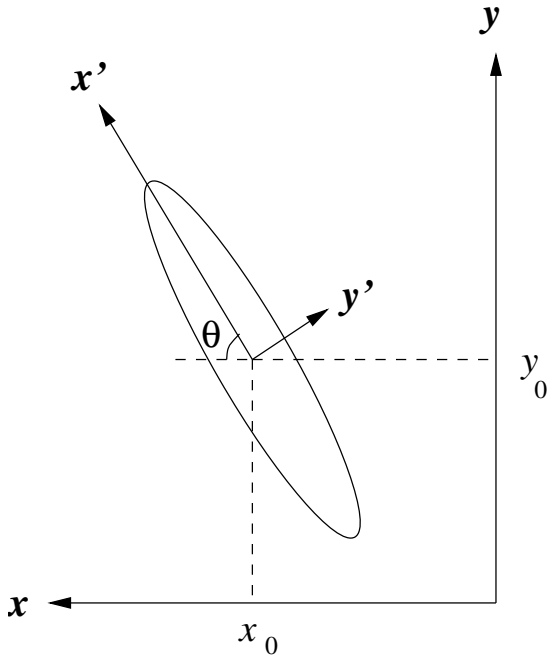}
 \caption{Reference systems. Both the $x'y'$- and the $xy$-coordinate systems are in the plane of the sky. The $x'y'$-system has its origin at the centre of the ellipse ($x_0,~y_0$), and $\theta$ is the angle between the $x$-axis and the $x'$-axis. The $x$ and $y$ axes are parallel to the RA and Dec axes.}
\end{figure}

\begin{figure*}
\includegraphics[width=150mm]{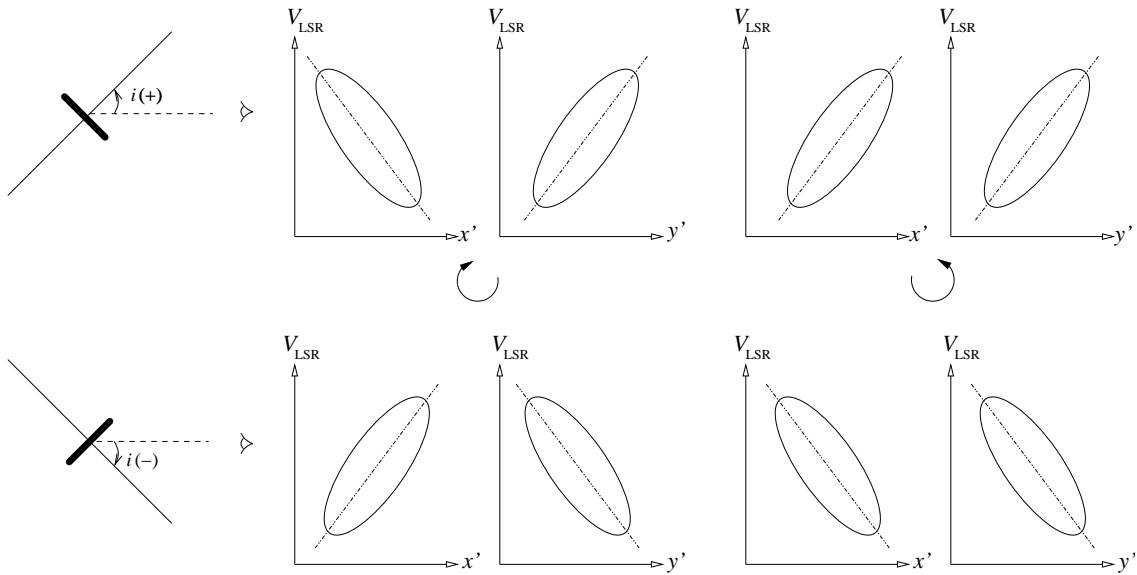}
 \caption{Position-velocity diagrams for a rotating and expanding ring model. The top panels correspond to a positive inclination angle and the bottom panels correspond to a negative inclination angle. 
The sense of rotation (clockwise or counterclockwise) as seen from the observer is indicated. The $x'$ and $y'$ axes are those defined in Fig.~1.}
\end{figure*}

Let $v_s$, $v_{\rmn{rot}}$, and $v_{\rmn{exp}}$ be the local standard of rest (LSR) systemic velocity, the rotation velocity, and the expansion velocity of the ring, respectively. Then the observed LSR velocity of a point in the ring can be expressed as
\begin{equation}
V_{\rmn{LSR}}=v_s+\frac{x'}{a}v_{\rmn{rot}}\sin i+\frac{y'}{a}v_{\rmn{exp}}\tan i.
\end{equation} 
Hence the observed $V_{\rmn{LSR}}$ will be a linear function of either $x'$ or $y'$, if only one type of motion (rotation or expansion, respectively) is present in the ring \citep{usc05}. 
Using equation (1), equation (4) can be written either in terms of the $x'$ or $y'$ coordinate as
\begin{equation}
\frac{[V_{\rmn{LSR}}-v_s-(x'/a)v_{\rmn{rot}}\sin i]^2}{(v_{\rmn{exp}}\sin i)^2}+\frac{x'^2}{a^2}=1,
\end{equation}
\begin{equation}
\frac{[V_{\rmn{LSR}}-v_s-(y'/a)v_{\rmn{exp}}\tan i]^2}{(v_{\rmn{rot}}\sin i)^2}+\frac{y'^2}{(a\cos i)^2}=1.
\end{equation}
Therefore, equations (5) and (6) indicate that the observed $V_{\rmn{LSR}}$ has a quadratic form (ellipse) expressed in terms of $x'$ or $y'$, when both motions are present.

In this model, we do not solve the radiative transfer through the ring, but we assume that the emission at a given $V_{\rmn{LSR}}$ comes from the point of the ring having this line-of-sight velocity component. We then use this information to construct position-velocity diagrams.

The orientation of the major axis of the ellipse in the position-velocity ($x'$-$V_{\rmn{LSR}}$ or $y'$-$V_{\rmn{LSR}}$) diagrams changes depending on the value of the inclination angle and on whether the sense of rotation is clockwise or counterclockwise as seen from the observer's point of view, as shown in Fig.~2. Accordingly, we are able to distinguish between a positive or negative value of the inclination angle and the sense of the rotation by doing a comparison between the position-velocity diagram delineated by the maser emission and the position-velocity diagrams expected for a rotating and expanding ring.
It is important to note that similar position-velocity diagrams can be obtained considering contraction instead of expansion in the ring, but changing the sign of the inclination angle and the sense of rotation. This ambiguity can be solved by constraining the value of the inclination angle (positive or negative) with additional information (see Section 3.2).

\subsection[]{Least-squares fit}
We carried out a least-squares fit of an ellipse to the observed emission, to estimate its spatial distribution on the sky.
In order to do this, we considered the curve on the
$xy$-plane given by equations (1)--(3) for a given set of values
of the parameters $(x_0,y_0)$ (the centre of the ellipse on the
plane of the sky), $a$ and $b$ (the semimajor and semiminor axes, respectively)
and $\theta$ (the angle between the $x$-axes of the two frames of reference).
We then compute the minimum distances
$d_j$ between the position $(x_j,y_j)$ of each of the masers and the
ellipse (these distances are measured along straight lines that pass through the maser and intersect the ellipse at right angles to
the curve). With these distances, we define the $\chi^2$ as
\begin{equation}
\chi^2={1\over {N-5}}\sum_{j=1}^N \left({d_j\over \sigma_j}\right)^2\,,
\label{chi2}
\end{equation}
where $N$ is the number of data points (i.e. the factor $N-5$ is the number of degrees of freedom), and $\sigma_j$ is the error associated with the position of the
maser spots. We then find the set of parameters $(x_0,y_0,a,b,\theta)$ (see
above) which give the minimum of $\chi^2$.

As a first approximation, we began by setting the values of $\sigma_j$ equal to the observational uncertainties $\Delta_j$ for the measured positions of the masers, and then finding the ellipse that gives the minimum value of ${\chi^2}$ 
(see equation \ref{chi2}). The actual structure of the maser-emitting region could be a ring of finite width, for which a broad ellipse is a rough approximation of its projection on the plane of the sky. Therefore, we do not expect maser spots to exactly trace an ellipse. We can characterise the width of the ring, assuming that the fitted ellipse traces the mean projected angular distance of the maser emission to the central star, and the actual emission will be distributed around this ellipse, with a dispersion $\Delta_e$. We treat $\Delta_e$ as a source of error for the ellipse fit, additional to the measured error, so that ${\sigma_j}^2={\Delta_j}^2+{\Delta_e}^2$. We then try different values of the width parameter ($\Delta_e$), until we obtain a fit with a minimum ${\chi^2}(\Delta_e)=1$. We consider $2 \Delta_e$ as the characteristic width of the maser ring. 

The $(x_0,y_0,a,b,\theta)$ parameters obtained from the 
minimization of ${\chi^2}$
give the best elliptical fit to the observed positions of the masers. 
With these spatial parameters, we carried out a kinematical fit, using the LSR velocities of the maser components
in order to define a ${\chi^2}_v$ of the form
\[
{\chi^2}_v={1\over {N-3}}\sum_{j=1}^N {1\over {{\sigma_v}^2}}
\]
\begin{equation}
\times\left(V_{\rmn{LSR}},j-v_s-{{x_j}'\over a}v_{\rmn{rot}}\sin i-
{{y_j}'\over a}v_{\rmn{exp}}\tan i\right)^2\,,
\end{equation}
where $N-3$ is the number of degrees of freedom, and $({x'}_j,{y'}_j)$ are given by equations (2)--(3).
Here $\sigma_v$ is the uncertainty in the observed LSR velocity that we adopt as the spectral resolution of the observations. 
The minimization of ${\chi^2}_v$ yielded the best values
for the systemic ($v_s$), rotation ($v_{\rmn{rot}}$), and expansion
($v_{\rmn{exp}}$) velocities.

\begin{table*}
 \centering
 \begin{minipage}{112mm}
  \caption{H$_2$O Masers in K3-35 (Region C)}
  \label{symbols}
  \begin{tabular}{@{}ccrcllccl}
  \hline
   $V_{\mathrm{LSR}}$ & \multicolumn{3}{c}{Flux Density} &
 \multicolumn{2}{c}{Position\footnote{ Units of right ascension are hours, minutes, and seconds, and units of declination are degrees, arcminutes, and arcseconds. Data from \citet{mir01} and \citet{gom03}.}}   &
  \multicolumn{3}{c}{Position Uncertainty\footnote{Relative position uncertainties (2$\sigma$) between maser spots.
The position of the 1.3~cm continuum emission peak is $\alpha(\rmn{J2000})=19^{\rmn{h}}27^{\rmn{m}}$44\fs0233, $\delta(\rmn{J2000})=21\degr30'$03\farcs441. The relative position uncertainty between the continuum and the H$_2$O masers is 0\farcs002.
The accuracy of the absolute positions is 0\farcs05.}} \\
   (km~s$^{-1})$ & \multicolumn{3}{c}{(mJy)}  &  $\alpha$(J2000) & $\delta$(J2000) & 
  \multicolumn{3}{c}{(arcsec)} \\
\hline
24.6 &  &   23 &  &   19 27 44.0243   & 21 30 03.438   &  &  &  0.010 \\  
24.0 &  &   61 &  &   19 27 44.0242   & 21 30 03.428   &  &  &  0.004 \\
23.3 &  &  218 &  &   19 27 44.02246  & 21 30 03.4460  &  &  &  0.0012 \\  
22.6 &  & 1010 &  &   19 27 44.022254 & 21 30 03.45146 &  &  &  0.00024 \\
22.0 &  & 1572 &  &   19 27 44.022364 & 21 30 03.45335 &  &  &  0.00014 \\
21.3 &  &  945 &  &   19 27 44.02247  & 21 30 03.4564  &  &  &  0.0003 \\  
20.7 &  &  201 &  &   19 27 44.02257  & 21 30 03.4625  &  &  &  0.0011 \\   
\hline
\end{tabular}
\end{minipage}
\end{table*}

\section{H$_2$O masers in K3-35 (region C)}
\subsection{Observational Data}
There are only two sets of VLA water maser observations towards the PN K3-35. The VLA 1999.7 epoch observations reported by \citet{mir01}, and the VLA 2002.3 epoch observations reported by \citet{deg04}.
In the latter observations, only a group of four maser spots was detected towards the central region of this source. No maser emission was detected at the tips of the bipolar lobes of the PN.

Note that in the \citet{deg04} paper the position of the continuum peak used to align the positions of the masers at the two epochs was not used with enough precision, resulting in a spurious shift of the maser spots from one epoch to another (see their Fig. 4). Using the position with the adequate precision, we find the positions of the masers at the two epochs to be consistent within the uncertainties, 0\farcs01 (2$\sigma$).

\subsection{Model Application and Results}
In our analysis, we have used the water maser data from the VLA 1999.7 epoch observations towards the central region of K3-35 reported by \citet{mir01}. 
At this epoch, the number of maser spots detected was larger than during the other epoch, allowing us a better identification of possible expansion and/or rotation motions at the centre of K3-35.
The velocity resolution of the VLA observations was 1.2~km~s$^{-1}$ and the accuracy in the relative positions of the water maser spots was of the order of milliarcseconds.
The positions of the observed water maser spots towards the core (region C) are listed in Table 1. We adopt as the origin of the $xy$-coordinate system the position of the 1.3 cm continuum emission peak.

Based on a least-squares fit to the positions of the maser spots (see Section 2.2) and using the observational uncertainties given in Table~1, we have found that the H$_2$O masers located towards the core of K3-35 can be fitted by a circular ring of $R\simeq0$\farcs021 radius ($\simeq$100~AU at the estimated distance of $\sim$5~kpc) with an angular width of 2$\Delta_e=0$\farcs003, observed at an inclination angle of $\vert i\vert\simeq\,55\degr$ (see Table 2 and Fig.~3).

\begin{figure}
 \includegraphics[width=84mm]{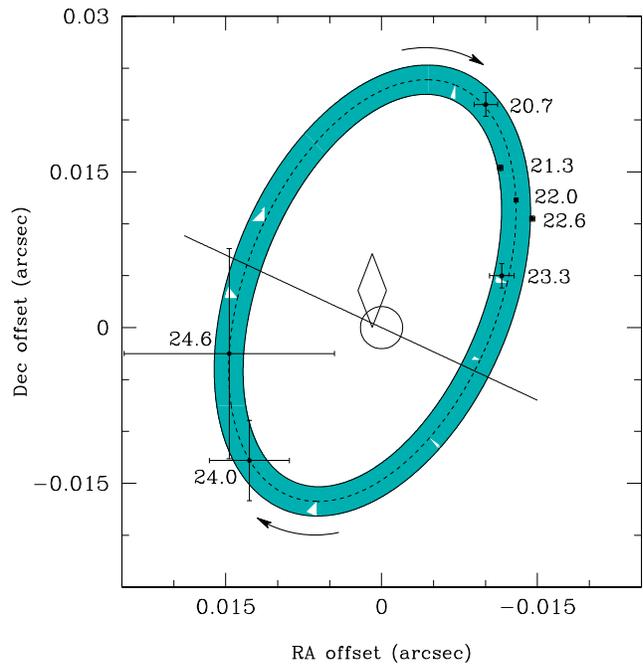}
 \caption{
Positions of the K3-35 water maser spots in offsets relative to the position of the 1.3~cm continuum emission peak \citep{mir01}. 
Each spot is labelled with its corresponding LSR velocity (in km s$^{-1}$).
The dashed ellipse corresponds to the least-squares fit to the maser spots positions whose parameters are indicated in Table 2. 
The open circle indicates the nominal position of the 1.3~cm continuum emission peak (see Table 1), its size is equal to the uncertainty of this position. The diamond indicates the position of the centre of the ellipse that was obtained from the fit.
The straight line shows the direction of the bipolar outflow traced by the innermost region of the jet. The arrows show the sense of rotation of the proposed ring (see Section 2.1).
The broad cyan ellipse has a width $2\Delta_e$, with $\Delta_e$ fulfilling $\chi^2(\Delta_e)=1$ (see Section 2.2).
The positional error bars indicate the uncertainties in the relative positions between maser spots given in Table 1.}
\end{figure}

Spectroscopic observations show that the northeastern lobe of the outflow is blueshifted, and the southwestern one is redshifted \citep{mir00}. If we assume that the ring traced by the water masers is perpendicular to the bipolar lobes then the inclination angle should be positive ($i\simeq+55\degr$). This means that the western half of the ring is closer to the observer.

The calculated position of the centre of the ellipse relative to the position of the 1.3 cm continuum emission peak $(x_0,y_0)$ is given in Table 2. Both positions are in agreement within the uncertainties (see the overlap of these positions in Fig.~3).

The kinematic trend is shown in Fig.~4. Since we have determined that the inclination angle of the ring is positive, the ambiguity between expansion and contraction can be solved when we compare the position-velocity diagrams delineated by the water masers (see the top panel of Fig. 4) and the position-velocity diagrams of the model (see Fig. 2). We have found that the ring traced by the masers rotates clockwise as seen from the observer at a velocity  $v_{\rmn{rot}}\simeq 3.1$~km~s$^{-1}$ and expands at a velocity $v_{\rmn{exp}}\simeq 1.4$~km~s$^{-1}$ (see Table 2).
The rotation and expansion velocities are estimated from a purely kinematical fit to the LSR velocities of the maser spots (see equation 8). 
The kinematical fit yields a ${\chi^2}_v$  value of 1.94, which means that the fit is good, assuming a conservative confidence level of 90\%.
Although not all maser spots may be completely independent, given the limited angular and spectral resolution, the observed velocity trend and the good kinematical fit suggest that these motions are real and systematic. We note that we have considered a single value of $v_{\rmn{rot}}$ instead a Keplerian rotation law. This is reasonable, since the velocity gradient over the ring width would be only $\sim$7\%. Tracing more subtle velocity variations would require more data points than the ones available.

The expansion velocity we found for the ring is close to that of thermal motions or subsonic turbulence. However we do not expect that thermal or turbulent motions could produce the systematic motions described in the position-velocity diagrams. The water masers seem to be tracing a spatio-kinematical structure with organised motions (at macroscopic scales).

\begin{figure}
 \includegraphics[width=84mm]{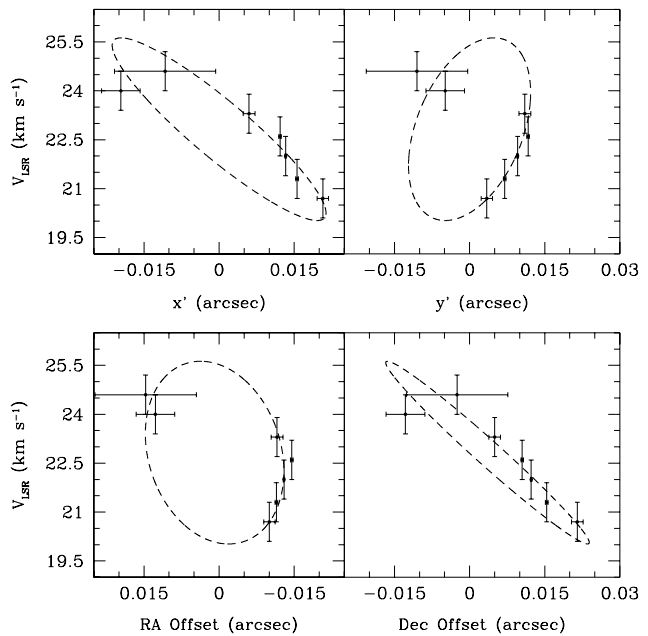}
 \caption{Position-velocity diagrams. \textit{Top}: The ordinate axis corresponds to the observed LSR velocity and the abscissa axis corresponds to the coordinate $x'$ or $y'$. \textit{Bottom}: Same as Fig.~4 (\textit{top}), but the abscissa axis corresponds to right ascension offset or declination offset relative to the position of the 1.3~cm continuum emission peak. 
The points correspond to the observed maser spots towards the centre of K3-35 and the dashed ellipses correspond to the kinematical model using the parameters listed in Table 2. The error bars in position are those shown in Fig.~3. The uncertainty in velocity is $\simeq$0.6~km~s$^{-1}$.}
\end{figure}

\section{Discussion}
From our model, we conclude that a ring is a likely explanation for the distribution and kinematics shown by the water masers located towards the centre of the PN K3-35.
This ring may be arising from the innermost region of a disc or torus probably formed at the end of the AGB phase. 
Since masers trace regions with very stringent physical conditions, it is not possible to know whether the ring is tracing part of a toroidal or a disc-like structure.
The kinematics of the ring in K3-35 suggests the presence of both rotating and expanding motions, as was also proposed in the young PN IRAS 17347-3139 \citep{deg04}. The estimated expansion and rotation velocity values for K3-35 are similar (a few km~s$^{-1}$) to those obtained from water maser observations of IRAS 17347-3139. 

The calculated expansion velocity of the ring ($\simeq$1.4~km~s$^{-1}$) in K3-35 is comparable to, but below, the expansion velocities of the tori inferred from interferometric CO observations in some PPNe and young PNe. For instance, the expansion velocities in M~1-92, M 2-9, M 2-56, and KjPn 8, have values in the range of $\simeq$5--8~km~s$^{-1}$ \citep{buj98,zwe97,cas02,for98}. It is noteworthy that the estimated expansion velocity in K3-35 obtained from water maser observations is quite similar to the slow expansion velocity ($\simeq$0.8~km~s$^{-1}$) deduced from CO observations in the PPN Red Rectangle \citep{buj05}.
However, the estimated sizes of the tori detected in CO observations in those sources are about one order of magnitude more than the size of structure traced by the water masers in K3-35, suggesting that the structures observed in CO are probably related to the outermost equatorial region, while the water masers could be arising from the innermost one, such as a circumstellar disc, closer to the central star.

\begin{table}
 \caption{Results of the rotating and expanding ring model
          fit to the H$_2$O masers towards the centre of K3-35.}
 \label{symbols}
 \begin{tabular}{@{}ccrcl}
  \hline
  Parameter & \multicolumn{4}{c}{Value$^a$} \\
  \hline
  $a$             &  & 0\farcs021   & $\pm$ & 0\farcs003    \\
  $b$             &  & 0\farcs012   & $\pm$ & 0\farcs002    \\
  $x_0\,^*$       &  & 0\farcs001   & $\pm$ & 0\farcs001    \\
  $y_0\,^*$       &  & 0\farcs004   & $\pm$ & 0\farcs004    \\
  PA              &  & 158$\degr$     & $\pm$ & 10$\degr$ \\
  $i$             &  & 55$\degr$      & $\pm$ & 7$\degr$ \\
  $v_s$           &  & 22.8    & $\pm$ & 0.5~km~s$^{-1}$ \\
  $v_{\rmn{exp}}$ &  & 1.4     & $\pm$ & 0.9~km~s$^{-1}$ \\
  $v_{\rmn{rot}}$ &  & 3.1     & $\pm$ & 0.8~km~s$^{-1}$ \\
  \hline
\end{tabular}
\medskip

$^a$ Note: Uncertainties are 2$\sigma$.\\
$^*$ Coordinates of the centre of the ellipse relative to the position of the 1.3~cm continuum emission peak.
\end{table}

In the case of K3-35, the water masers may be delineating an ellipse on the plane of the sky with a semimajor axis of $\simeq$100 AU and a PA$\simeq$158$\pm$10\degr, suggesting the presence of a disc or torus projected on the sky. The major axis of the ellipse is almost perpendicular (within the uncertainties) to the direction of the bipolar emission traced by the innermost region of the jet, with a PA$\simeq$65\degr observed by \citet{mir01}, suggesting that the disc or torus could be physically related to the collimation of the bipolar outflow.

Recently, \citet{hug07} found that jets typically appear a few hundred years after the torus formation in PPNe and young PNe.
This time sequence provides evidence that jets and tori are physically related. In the case of K3-35, \citet{vel07} modelled its radio continuum emission as a precessing dense jet with an age of $\sim$40~yrs. On the other hand,
our estimate of the dynamical age of the ring traced by the maser emission is about $\sim$350 years, using its radius ($\sim$100~AU) and assuming a uniform expansion velocity of $\sim$1.4~km~s$^{-1}$. 
These values would, in principle, suggest a lag in time similar to the value found by \citet{hug07}. However, the dynamical age we derive is not reliable enough as an age estimate. Proper motion studies of the water maser emission (e.g. using e-MERLIN and VLBA) could provide better estimates.

The kinematics of the ring in K3-35 suggests the presence of both expansion and rotation. 
A rough estimate for the central stellar mass can be obtained by assuming that the total energy (kinetic plus gravitational)
of the masing gas is close to zero. 
In this case, $M\simeq (R/2G)(v_{\rmn{exp}}^2+v_{\rmn{rot}}^2)$, where $R\simeq100~(D/5~\rmn{kpc})$~AU, is the radius of the ring. 
Hence, $M\simeq[0.7~(D/5~\rmn{kpc})\pm 0.3]~\rmn{M}_{\sun}$, where $D$ is the source distance. The error in the mass only includes the errors in the fitted parameters. This estimate of the central mass is in agreement with the core mass required for a PNe, according to evolutionary models \citep{kwo03}.

\section*{Acknowledgments}
LU and YG acknowledge support from DGAPA-UNAM grant IN100407 and CONACyT grant 49947. 
LU is supported by Secretar\'{\i}a de Estado de Universidades e Investigaci\'on of MEC. 
AR and JC are supported by CONACyT grants 46828-F and 61547. 
LFM acknowledges support from MEC AYA2005-01495 grant (co-funded with FEDER funds). 
GA, JFG, and JMT are supported by the MEC AYA2005-05823-C03 grant (co-funded with FEDER funds). GA, JFG, LFM, JMT and LU are also supported by Consejer\'{\i}a de Innovaci\'on, Ciencia y Empresa of Junta de Andaluc\'{\i}a. 
We thank L. F. Rodr\'{\i}guez for valuable comments.
We are thankful to our referee, Anita Richards, for her useful comments on the manuscript.

\bsp

\label{lastpage}


\begin{thebibliography}{99}
\bibitem[\protect\citeauthoryear{Balick, Preston \& Icke}{1987}]{bal87} Balick B., Preston H. L., Icke V., 1987, AJ, 94, 1641
\bibitem[\protect\citeauthoryear{Bieging \& Nguyen-Quang-Rieu}{1988}]{bie88} Bieging J. H., Nguyen-Quang-Rieu, 1988, ApJ, 324, 516
\bibitem[\protect\citeauthoryear{Bujarrabal, Alcolea \& Neri}{1998}]{buj98} Bujarrabal V., Alcolea J., Neri R., 1998, ApJ, 504, 915
\bibitem[\protect\citeauthoryear{Bujarrabal et al.}{2005}]{buj05} Bujarrabal V., Castro-Carrizo A., Alcolea J., Neri R., 2005, A\&A, 441, 1031
\bibitem[\protect\citeauthoryear{Bujarrabal et al.}{2003}]{buj03} Bujarrabal V., Neri R., Alcolea J., Kahane C., 2003, A\&A, 409, 573
\bibitem[\protect\citeauthoryear{Castro-Carrizo et al.}{2002}]{cas02} Castro-Carrizo A., Bujarrabal V., S\'anchez Contreras C., Alcolea J., Neri R., 2002, A\&A, 386, 633
\bibitem[\protect\citeauthoryear{Chevalier \& Luo}{1994}]{che94} Chevalier R. A., Luo D., 1994, ApJ, 421, 225
\bibitem[\protect\citeauthoryear{de Gregorio-Monsalvo et al.}{2004}]{deg04} de Gregorio-Monsalvo I., G\'omez Y., Anglada G., Cesaroni R., Miranda L. F., G\'omez J. F., Torrelles J. M., 2004, ApJ, 601, 921 
\bibitem[\protect\citeauthoryear{Feibelman}{1985}]{fei85} Feibelman W. A., 1985, AJ, 90, 2550
\bibitem[\protect\citeauthoryear{Forveille et al.}{1998}]{for98} Forveille T., Huggins P. J., Bachiller R., Cox P., 1998, ApJ, 495, L111
\bibitem[\protect\citeauthoryear{Garc\'{\i}a-Segura}{1997}]{gar97} Garc\'{\i}a-Segura G., 1997, ApJ, 489, L189
\bibitem[\protect\citeauthoryear{G\'omez et al.}{2005a}]{gom05a} G\'omez J.~F., de Gregorio-Monsalvo I., Lovell J.~E.~J., Anglada G., Miranda L.~F., Su\'arez O., Torrelles J.~M., G\'omez Y., 2005a, MNRAS, 364, 738 
\bibitem[\protect\citeauthoryear{G\'omez et al.}{2003}]{gom03} G\'omez Y., Miranda L. F., Anglada G., Torrelles J. M., 2003 in Kwok S., Dopita M., Sutherland R., eds, Proc. IAU Symp. 209, Planetary Nebulae: Their Evolution and Role in the Universe. Astron. Soc. Pac., p. 263
\bibitem[\protect\citeauthoryear{G\'omez et al.}{2005b}]{gom05b} G\'omez Y., Tafoya D., Anglada G., Franco-Hern\'andez R., Torrelles J. M., Miranda L. F., 2005b, Mem. Soc. Astron. Italiana, 76, 472
\bibitem[\protect\citeauthoryear{Huggins}{2007}]{hug07} Huggins P. J., 2007, ApJ, 663, 342
\bibitem[\protect\citeauthoryear{Icke}{2003}]{ick03} Icke V., 2003, A\&A, 405, L11
\bibitem[\protect\citeauthoryear{Imai et al.}{2002}]{ima02} Imai H., Obara K., Diamond P. J., Omodaka T., Sasao T., 2002, Nat, 417, 829
\bibitem[\protect\citeauthoryear{Kwok}{2003}]{kwo03} Kwok S., 2003 in Kwok S., Dopita M., Sutherland R., eds, Proc. IAU Symp. 209, Planetary Nebulae: Their Evolution and Role in the Universe. Astron. Soc. Pac., p. 3
\bibitem[\protect\citeauthoryear{Livio \& Soker}{1988}]{liv88} Livio M., Soker N., 1988, ApJ, 329, 764
\bibitem[\protect\citeauthoryear{Mellema}{1995}]{mel95} Mellema G., 1995, MNRAS, 277, 173
\bibitem[\protect\citeauthoryear{Miranda et al.}{2000}]{mir00} Miranda L. F., Fern\'andez M., Alcal\'a J. M., Guerrero M. A., Anglada G., G\'omez Y., Torrelles J. M., Aaquist O. B., 2000, MNRAS, 311, 748
\bibitem[\protect\citeauthoryear{Miranda et al.}{2001}]{mir01} Miranda L. F., G\'omez Y., Anglada G., Torrelles J. M., 2001, Nat, 414, 284
\bibitem[\protect\citeauthoryear{Morris}{1987}]{mor87} Morris M., 1987, PASP, 99, 1115
\bibitem[\protect\citeauthoryear{Pascoli}{1985}]{pas85} Pascoli G., 1985, A\&A, 147, 257
\bibitem[\protect\citeauthoryear{Phillips}{2004}]{phi04} Phillips J.~P., 2004, MNRAS, 353, 589 
\bibitem[\protect\citeauthoryear{Reyes-Ruiz \& L\'opez}{1999}]{rey99} Reyes-Ruiz M., L\'opez J. A., 1999, ApJ, 524, 952
\bibitem[\protect\citeauthoryear{Riera et al.}{2003}]{rie03} Riera A., Garc\'{\i}a-Lario P., Manchado A., Bobrowsky M., Estalella R., 2003, A\&A, 401, 1039
\bibitem[\protect\citeauthoryear{Rijkhorst, Mellema \& Icke}{2005}]{rij05} Rijkhorst E.-J., Mellema G., Icke, V., 2005, A\&A, 444, 849
\bibitem[\protect\citeauthoryear{Sabin, Zijlstra \& Greaves}{2007}]{sab07} Sabin L., Zijlstra A.~A., Greaves J.~S., 2007, MNRAS, 376, 378 
\bibitem[\protect\citeauthoryear{Sahai \& Trauger}{1998}]{sah98a} Sahai R., Trauger J. T., 1998, AJ, 116, 1357
\bibitem[\protect\citeauthoryear{Sahai et al.}{1998}]{sah98b} Sahai R., Hines D. C., Kastner J. H., Weintraub D. A., Trauger J. T., Rieke M. J., Thompson R. I., Schneider G., 1998, ApJ, 492, L163
\bibitem[\protect\citeauthoryear{Soker}{2006}]{sok06} Soker N., 2006, ApJ, 645, L57
\bibitem[\protect\citeauthoryear{Soker \& Livio}{1994}]{sok94} Soker N., Livio M., 1994, ApJ, 421, 219
\bibitem[\protect\citeauthoryear{Soker \& Rappaport}{2000}]{sok00} Soker N., Rappaport S., 2000, ApJ, 538, 241
\bibitem[\protect\citeauthoryear{Taam \&  Bodenheimer}{1989}]{taa89} Taam R. E., Bodenheimer P., 1989, ApJ, 337, 849
\bibitem[\protect\citeauthoryear{Uscanga et al.}{2005}]{usc05} Uscanga L., Cant\'o J., Curiel S., Anglada G., Torrelles J. M., Patel N. A., G\'omez J. F., Raga A. C., 2005, ApJ, 634, 468
\bibitem[\protect\citeauthoryear{Vel\'azquez et al.}{2007}]{vel07} Vel\'azquez P. F., G\'omez Y., Esquivel A., Raga A. C., 2007, MNRAS, 382, 1965
\bibitem[\protect\citeauthoryear{Vlemmings, Diamond \& Imai}{2006}]{vle06} Vlemmings W.~H.~T., Diamond P.~J., Imai H., 2006, Nat, 440, 58 
\bibitem[\protect\citeauthoryear{Zhang}{1995}]{zha95}Zhang C. Y., 1995, ApJS, 98, 659
\bibitem[\protect\citeauthoryear{Zweigle et al.}{1997}]{zwe97}Zweigle J., Neri R., Bachiller R., Bujarrabal V., Grewing M., 1997, A\&A, 324, 624
\end{thebibliography}
\end{document}